\documentclass[%
 reprint,
groupedaddress,
 amsmath,amssymb,
 aps,
 pra,
]{revtex4-2}

\usepackage{dcolumn}
\usepackage{bm}
\usepackage{mathtools} 
\usepackage{color}  
\usepackage{float}
\usepackage{braket}
\usepackage{mhchem}

\definecolor{mscolor}{rgb}{0,0.5,0.5}

\definecolor{AkScolor}{rgb}{0.4,0,0}

\definecolor{tgcolor}{rgb}{0.5,0,0.5}

{}
\definecolor{phcolor}{rgb}{0.5,0,0.5}

\newcommand{\UWM}{Department of Physics, University of Wisconsin-Madison, 1150 University Avenue, Madison, WI, 53706, USA}

\newcommand {\rsub}[1]{\textcolor{black}{#1}}

\begin{document}

\title{Robust atom-photon gate for quantum information processing}

\author{Omar Nagib}
 \email{onagib@wisc.edu}
\affiliation{\UWM}

\author{P. Huft}
\affiliation{\UWM}

\author{A. Safari}
\affiliation{\UWM}

\author{M. Saffman}
\affiliation{\UWM}

\date{\today}

\begin{abstract}

We propose a scheme for two-qubit gates between a flying photon and an atom in a cavity. The atom-photon gate setup consists of a cavity and a Mach-Zehnder interferometer with doubly degenerate ground and excited state energy levels mediating the atom-light interaction.  We provide an error analysis of the gate and model important errors, including spatial mode mismatch between the photon and the cavity, spontaneous emission, cavity losses, detunings, and random fluctuations of the cavity parameters and frequencies. Error analysis shows that the gate protocol is more robust against experimental errors compared to previous atom-photon gates and achieves higher fidelity.

\end{abstract}

\maketitle


\section{Introduction}
Interconnecting multiple quantum processors or sensors in a quantum network\cite{Kimble2008} is a promising approach to achieve distributed quantum computing based on a modular architecture, and enhanced quantum sensing \cite{Monroe2014, Nadlinger2022, Zhang2022}. Furthermore, quantum networking enables quantum teleportation and secure quantum communication over large distances \cite{Olmschenk2009, Langenfeld2021, Nichol2022,Krutyanskiy2023b}.  Despite the significant progress in the field, building a quantum network with high entanglement fidelity and rate remains an outstanding challenge \cite{Pompili2021,Covey2023, Young2022}. Such a network relies on  entangling gates between stationary and flying qubits to distribute entanglement among remote quantum processors and sensors.

Twenty years ago in a pioneering work, Duan and Kimble proposed the use of photon-atom interactions in a cavity to realize entangling operations between atomic and photonic qubits and between pairs of photonic qubits \cite{Duan2004, Duan2005} (see also \cite{Hoffmann2003}). This provided a new approach to remote entanglement and opened the avenue for various applications in quantum information processing, including deterministic entanglement between atoms and photons, nondestructive Bell state measurements, nondestructive photon measurement, photon-photon entanglement, nonlocal entangling gates between remote atoms, quantum teleportation by photons, and generation of cat states, to name a few \cite{Reiserer2014,Reiserer2015, Daiss2021,Welte2018, Reiserer2013,Langenfeld2021, Welte2021, Hacker2019, Cohen2018}.

To realize atom-photon gates, traditional schemes use atoms with three energy levels: two non-degenerate ground states, and an excited state, which interact with right or left hand circularly polarized photons \cite{Duan2004, Duan2005, Reiserer2014}. Since the first experimental realization of atom-photon gates to this day, a major challenge has been to implement these gates with high fidelity. The fidelity of atom-photon gates or applications thereof (e.g., remote atom-atom entanglement, Bell state measurements, teleportation, entangling two atoms in a cavity) have been typically limited to the range of $75-80 \%$ (with the exception of teleportation with fidelity in the range of $85-90\%$) \cite{Reiserer2014,Daiss2021,Welte2018,Langenfeld2021, Welte2021}. Many sources of error have been identified as contributing to the infidelity, e.g., spatial mode mismatch between the photon and the cavity mode, multiphoton effects due to using coherent light sources, frequency fluctuations, and cavity losses \cite{Reiserer2014, Daiss2021,Welte2018, Reiserer2013,Langenfeld2021, Welte2021,Hacker2019}.

It was previously proposed, in the context of atoms coupled to waveguides, that using an energy scheme with two degenerate ground states and excited states allows for atom-photon gates that are more robust to errors \cite{YLi2012}. In this work, we adopt such a scheme for atoms in a cavity interacting with photons. We develop an error model that takes into account spatial mode mismatch between the photon and the cavity, spontaneous emission, cavity losses, finite cooperativity, and photon-cavity/atom-cavity frequency detunings. Error analysis shows that the new scheme can achieve higher average fidelity and is less sensitive to the errors just mentioned. In Sec. \ref{Atom-photon two-qubit gate}, we describe the energy scheme of the atom-photon $\sf CZ$ gate and propose how to realize it in atom-cavity systems. Section \ref{Error analysis} is the result of our error analysis, where we compare the fidelity and success probability in our scheme against previous ones. Section \ref{Discussion} concludes with a  discussion.

\section{Atom-photon two-qubit gate}\label{Atom-photon two-qubit gate}

We will use a scheme that implements a two-qubit $\sf CZ$ gate between atomic and photonic qubits by scattering a photon off the atom-cavity system. One side of the cavity is perfectly reflective while the other side is partially reflective. We will use the energy scheme and the experimental setup proposed in Ref. \cite{YLi2012} but adopt it for an atom that is confined in a cavity \cite{Duan2004, Duan2005, An2009, QChen2009, GYWang2016}. The relevant energy levels are shown in Fig. \ref{energy_scheme}. We have two degenerate ground states, $\{\ket{g_+},\ket{g_{-}} \}$, and two degenerate excited states, $\{\ket{e_+},\ket{e_{-}} \}$. $\ket{g_+}$ only couples to $\ket{e_+}$ via a $\ket{\sigma_{+}}$ photon while $\ket{g_{-}}$ only couples to $\ket{e_{-}}$ via a $\ket{\sigma_{-}}$ photon. By scattering a photon off a cavity with an atom inside we can achieve, as we show later, the following relations: 

\begin{figure}[!t] 
    \centering
    \includegraphics[width=0.5\textwidth]{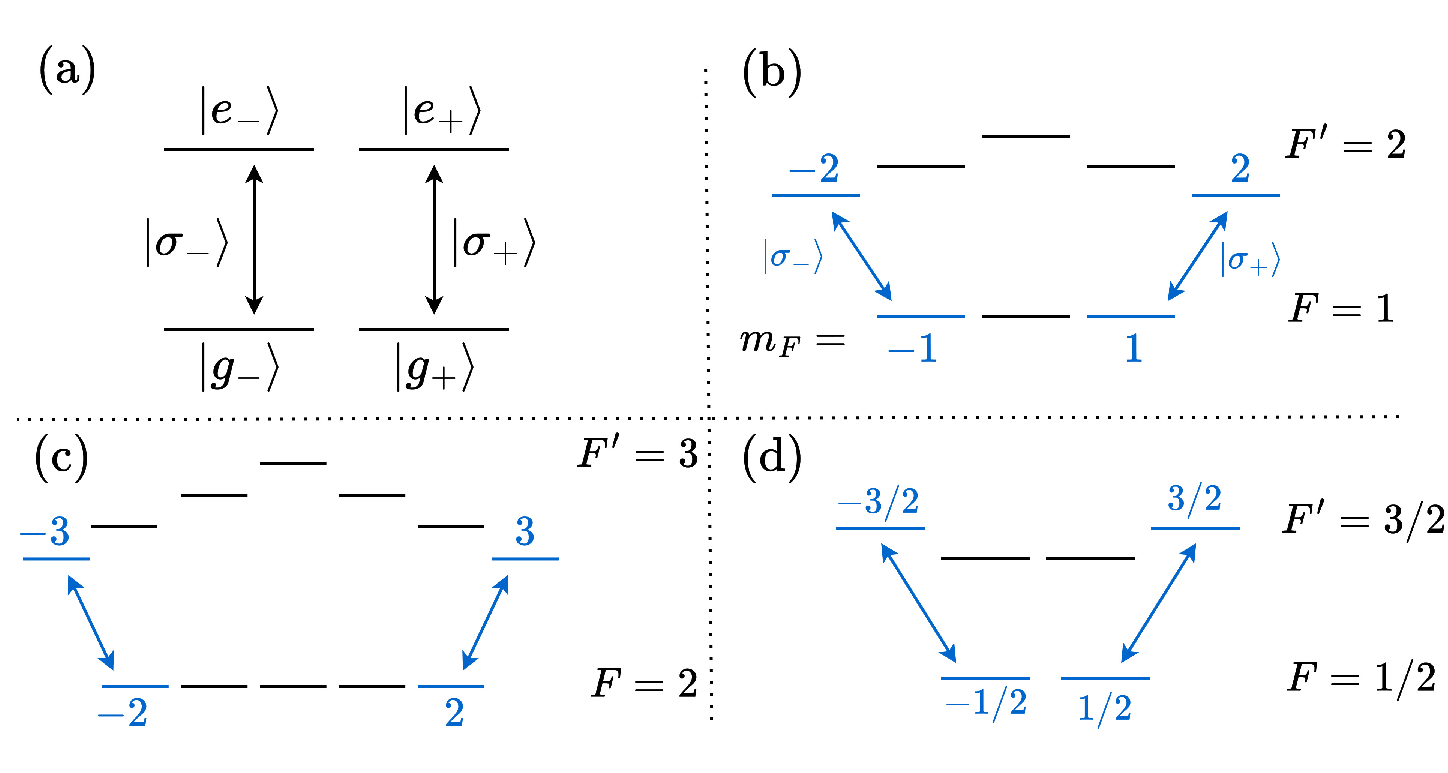}
    \caption{Energy levels needed for an entangling gate between photonic and atomic qubits. (a) The ground and excited states are doubly degenerate. Three possible candidate energy levels are shown in (b), (c), and (d). The energy level scheme can be realized using AC Stark shift. The energy levels corresponding to $\ket{g_{\pm}}$ and $\ket{e_{\pm}}$ are in  blue.}
    \label{energy_scheme}
\end{figure}

\begin{subequations}
\begin{eqnarray}\label{scatter_on}
\ket{\sigma_{\pm}}\ket{g_{\pm}} &\rightarrow& \ket{\sigma_{\pm}}\ket{g_{\pm}}, \\ 
\label{scatter_off}
\ket{\sigma_{\mp}}\ket{g_{\pm}} &\rightarrow& -\ket{\sigma_{\mp}}\ket{g_{\pm}},
\end{eqnarray}
\label{eq.inout}
\end{subequations}
i.e., there is no phase shift when the photon and the atom-cavity system are coupled, and a phase shift of $\pi$ when they are uncoupled.

It follows from the above relations that if we scatter a horizontally polarized photon, $\ket{H}=(\ket{\sigma_{+}}+\ket{\sigma_{-}})/\sqrt{2}$, off the cavity, then we will get

\begin{equation}\label{Z_gate}
\ket{H}\ket{g_{\pm}} \rightarrow \pm \ket{V}\ket{g_{\pm}}
\end{equation} 
where $\ket{V}=(\ket{\sigma_{+}}-\ket{\sigma_{-}})/\sqrt{2}$ is vertical polarization. If we denote $\ket{g_{+}}=\ket{0}^{\rm a}$ and $\ket{g_{-}}=\ket{1}^{\rm a}$ (the superscript $\rm a$ stands for atomic qubit), then Eq. (\ref{Z_gate}) represents a $\sf Z$ gate on the atomic qubit $\ket{\phi}^a$:
\begin{equation}\label{Zgate}
\ket{H}\ket{\phi}^{\rm a}\rightarrow {\sf Z}_{\rm a} \ket{V}\ket{\phi}^{\rm a}.
\end{equation} 
Note that the photon polarization is flipped during the scattering.

\begin{figure}[!t]
    \centering
    \includegraphics[width=0.5\textwidth]{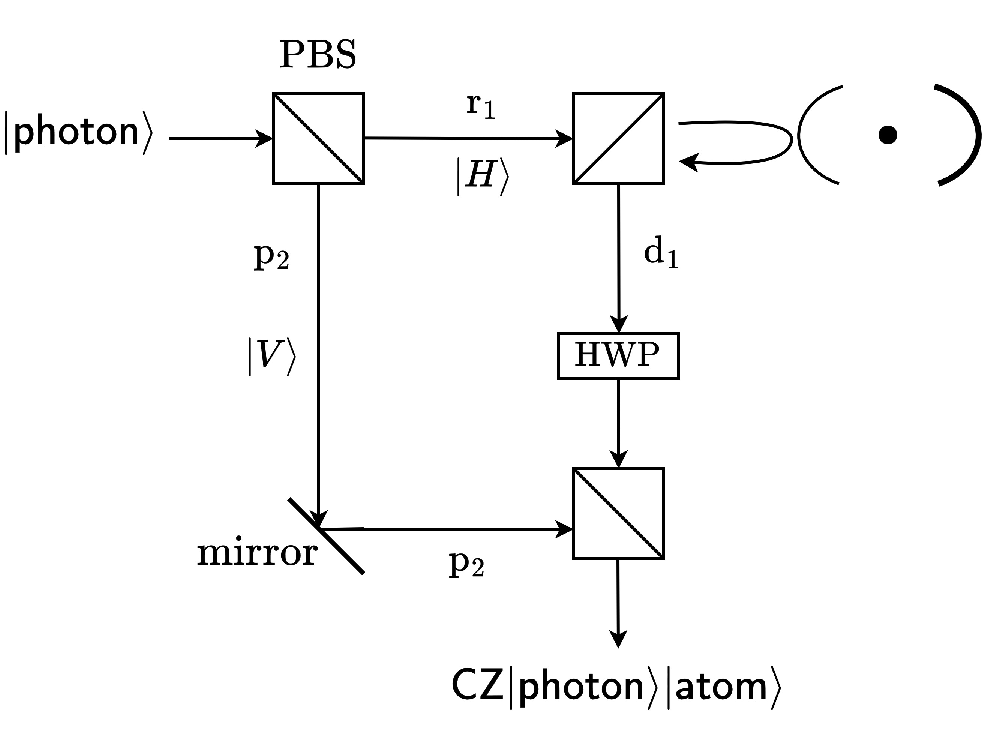}
    \caption{$\sf CZ$ atom-photon gate setup. $\ket{V}=\ket{0}^{\rm p}$ does nothing on the atomic qubit as it does not scatter off the cavity. $\ket{H}=\ket{1}^{\rm p}$ does a $\sf Z$ gate by scattering off the cavity. The combined operation gives $\sf CZ$. All the three beam splitters here are polarizing beam splitters.}
    \label{CZ_gate}
\end{figure}

The physical implementation of the atom-photon $\sf CZ$ gate is shown in Fig. \ref{CZ_gate}. The setup corresponds to a Mach–Zehnder interferometer (MZI). The first polarizing beam splitter (PBS) separates the $\ket{H}$ and $\ket{V}$ components into two paths, such that $\ket{V}=\ket{0}^{\rm p}$ does nothing to the atomic qubit (because it does not interact with the cavity), while $\ket{H}=\ket{1}^{\rm p}$ implements a $\sf Z$ gate on the atom by scattering (the superscript $\rm p$ denotes photonic qubit). After $\ket{H}$ is reflected off the cavity, its polarization flips to $\ket{V}$ and is reflected down by the PBS. A half-waveplate (HWP) is inserted to restore the polarization back to $\ket{H}$. The third PBS combines the two polarization components into a single path, and we end up with a $\sf CZ$ operation between the atom and the photon.

To realize the required energy scheme of Fig. \ref{energy_scheme}, we propose to use AC Stark shifts, which can be realized in a linearly polarized optical dipole trap. Several possible candidate energy levels are shown in Fig. \ref{energy_scheme}. The ground state energy levels $F$ all experience the same value of AC Stark shift while the excited states $F'$ experience a shift that depends on the Zeeman state. More precisely the requirement is that the ground state only has a scalar shift and that the excited state has tensor but no, or negligible, vector shift. The right-most and left-most energy levels of $F$ and $F'$ (in blue) correspond to $\ket{g_{\pm}}$ and $\ket{e_{\pm}}$. This structure ensures that $\ket{\sigma_{+}}$ and $\ket{\sigma_{-}}$ have the same frequency, and that $\ket{g_{\pm}}$ only couple to $\ket{e_{\pm}}$ via $\ket{\sigma_{\pm}}$. This energy scheme can be realized in many atoms. As an example, the energy level structure of Fig. \ref{energy_scheme}(c) was previously experimentally realized in $^{87}\ce{Rb}$ \cite{Reiserer2014}. Another candidate for Fig. \ref{energy_scheme}(d) is $^{171}\ce{Yb}$ \cite{Ma2022}. Details of the implementation for $^{87}\rm Rb$ are discussed in Appendix \ref{Rb_implement}.

We proceed to describe how Eqs. \eqref{scatter_on} and \eqref{scatter_off} are realized by scattering photons off a cavity \cite{Duan2004, Duan2005, An2009, QChen2009, GYWang2016}. First, consider the energy levels of an atom in Fig. \ref{energy_scheme}(a) that interact with a $\ket{\sigma_{\pm}}$ photon scattering off the cavity. Here we assume that the photon frequency is $\omega_{\rm p}$ and the frequency between the two atomic energy levels $\ket{g_{\pm}}$ and $\ket{e_{\pm}}$ is $\omega_{\rm a}$. Using the Jaynes-Cummings model and input-output theory, general expressions for the reflection coefficient that the scattered photon experiences, and which are valid in both  the strong and  weak coupling regimes of the cavity were derived in \cite{An2009, QChen2009, Reiserer2015}:

\begin{eqnarray}\label{rn2}
r_{\rm c}&=&\dfrac{(i\Delta_{\rm c}-1)(i\Delta_{\rm a}+1)+2C}{(i\Delta_{\rm c}+1)(i\Delta_{\rm a}+1)+2C},\\
\label{ro2}
r_{\rm nc}&=&\dfrac{i\Delta_{\rm c}-1}{i\Delta_{\rm c}+1}.
\end{eqnarray}
Here $r_{\rm c}$ is the reflection coefficient when the photon is coupled to the atom-cavity system, i.e., $\ket{\sigma_{\pm}}\ket{g_{\pm}}$, $r_{\rm nc}$ is the reflection coefficient when the photon is not coupled, i.e., $\ket{\sigma_{\pm}}\ket{g_{\mp}}$, $\Delta_{\rm c}=(\omega_{\rm p}-\omega_{\rm c})/\kappa$, $\Delta_{\rm a}=(\omega_{\rm p}-\omega_{\rm a})/\gamma$ are the fractional detunings,  $\omega_{\rm c}$ is the cavity frequency, $C=g^2/ 2\gamma \kappa$ is the the cooperativity,  $g$ is the atom-cavity coupling constant, $\kappa$ is the cavity decay rate, and $\gamma$ is the atom decay rate. These expressions are valid for sufficiently large $\kappa$ such that there is only weak atomic excitation from the single photon pulse \cite{An2009, QChen2009}. Here we use the expressions taken from Ref. \cite{Reiserer2015} and  $\kappa$ and $\gamma$ are half-width at half-maximum (HWHM) quantities.

Thus we get the following scattering relations

\begin{eqnarray}
\label{scatter_on_rn}
\ket{\sigma_{\pm}}\ket{g_{\pm}}& \rightarrow& r_{\rm c} \ket{\sigma_{\pm}}\ket{g_{\pm}} \\
\label{scatter_off_ro}
\ket{\sigma_{\mp}}\ket{g_{\pm}} &\rightarrow& r_{\rm nc} \ket{\sigma_{\mp}}\ket{g_{\pm}}
\end{eqnarray} 

In the limit of large $C \gg 1$ and small $\Delta_{\rm c/a} \ll 1$, we have $r_{\rm c} \approx 1$, $r_{\rm nc}\approx-1$, and we recover the desired relations Eqs. \eqref{eq.inout}. 

We proceed to describe the essential feature of this atom-photon gate that makes it robust to errors. Consider a horizontally polarized photon $\ket{H}$ impinging on the atom-cavity state $\alpha  \ket{0}^{\rm a}+\beta \ket{1}^{\rm a}$ from the input port on the left (see Fig. \ref{CZ_gate}). Ideally, if $r_{\rm c} =1$ and $r_{\rm nc}=-1$, the photon performs a $\sf Z$ gate on the atomic qubit, its polarization  flips to $\ket{V}$, and then it will exit the system through the PBS on the bottom right (output port). Detecting the photon heralds a successful $\sf Z$ operation on the atomic qubit. However, if the cavity is not ideal, i.e.,  $r_{\rm c}  \neq - r_{\rm nc}$, then the quantum state immediately after scattering becomes (see Appendix \ref{CZ_error_new}):

\begin{equation}
(r_{\rm c}+r_{\rm nc})\ket{H}(\alpha \ket{0}^{\rm a}+\beta \ket{1}^{\rm a})+(r_{\rm c}-r_{\rm nc})\ket{V}(\alpha \ket{0}^{\rm a}-\beta \ket{1}^{\rm a})
\end{equation} 

The first term is the error term where the photon failed to perform a $\sf Z$ gate and its polarization remained the same (which occurs with probability proportional to $|r_{\rm c}+r_{\rm nc}|^2$). The second term is where the photon successfully performed a $\sf Z$ gate and its polarization flipped (which occurs with probability proportional to $|r_{\rm c}-r_{\rm nc}|^2$). The error component (i.e., the horizontally polarized photon) is reflected back and exits through the input port it originally came from. Therefore, the error component is never detected at the output port (bottom right). Only the second term with vertical polarization and successful $\sf Z$ operation is reflected down and is detected through the output port. Therefore, any failure to perform the $\sf Z$ gate (due to errors) is rejected out of the system and is never detected at the output port, i.e., the photon is lost instead of affecting the fidelity. This same mechanism allows this gate to reduce other error sources like spatial mode mismatch between the photon and the cavity. 

We would like to point out another feature of our scheme. Because of the symmetric energy level structure used, the fidelity of the atom-photon $\sf CZ$ gate does not depend on the initial atomic state $\alpha\ket{0}^{\rm a}+\beta\ket{1}^{\rm a}$, i.e., the fidelity does not depend on $\alpha$ and $\beta$ even in the presence of the errors (see the formula for the fidelity in Appendix \ref{CZ_error_new}). This is a desirable feature in quantum information processing, e.g., in quantum teleportation of qubits. This is in contrast to earlier schemes \rsub{\cite{Duan2004, Duan2005,Reiserer2014}}, where the gate fidelity is higher for certain initial atomic qubit states and lower for others (see the formula for the fidelity in Appendix \ref{CZ_error_old}). 

\begin{figure}[!t] 
    \centering
    \includegraphics[width=0.5\textwidth]{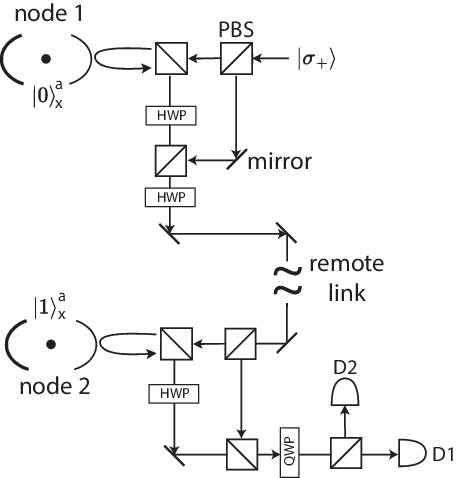}
    \caption{A setup to generate remote atom-atom entanglement between nodes using the proposed atom-photon gate. A $\ket{\sigma_+}$ photon from the outside hits the two cavities successively performing two $\sf CZ$ gates. Measurement of the photon polarization heralds the generation of atom-atom entanglement. All beamsplitters are polarizing beam splitters.}
    \label{atom_atom}
\end{figure}

The atom-photon gate can be used to generate remote atom-atom entanglement between two cavities as shown in Fig. \ref{atom_atom}. The two interferometers next to the two cavities perform atom-photon $\sf CZ$ gates. The atom in node 1  is initially prepared in $\ket{0}_x^{\rm a}=(\ket{0}^{\rm a}+\ket{1}^{\rm a})/\sqrt2$ while the one in node 2 is in $\ket{1}_x^{\rm a}=(\ket{0}^{\rm a}-\ket{1}^{\rm a})/\sqrt2$. We then  input  a $\ket{\sigma_+}$ photon, perform an atom-photon $\sf CZ$ gate at node 1, and use the output photon to perform  an atom-photon  $\sf CZ$ gate at node 2. After the output photon passes through a quarter waveplate (QWP), we measure its polarization. Ideally, detection of the photon polarization heralds the generation of the maximally entangled atom-atom Bell states according to (see Appendix \ref{CZ_error_new}):
\begin{subequations}
\begin{eqnarray} 
\ket{H}: \frac{\ket{01}^{\rm a} +\ket{10}^{\rm a}}{\sqrt2}\\
\ket{V}: \frac{\ket{00}^{\rm a} +\ket{11}^{\rm a}}{\sqrt2},
\end{eqnarray}\end{subequations}
where $\ket{H}$ and $\ket{V}$ correspond to clicks on detectors D1 and D2.

\section{Error analysis} \label{Error analysis}

\begin{figure}[!t] 
    \centering
    \includegraphics[width=0.5\textwidth]{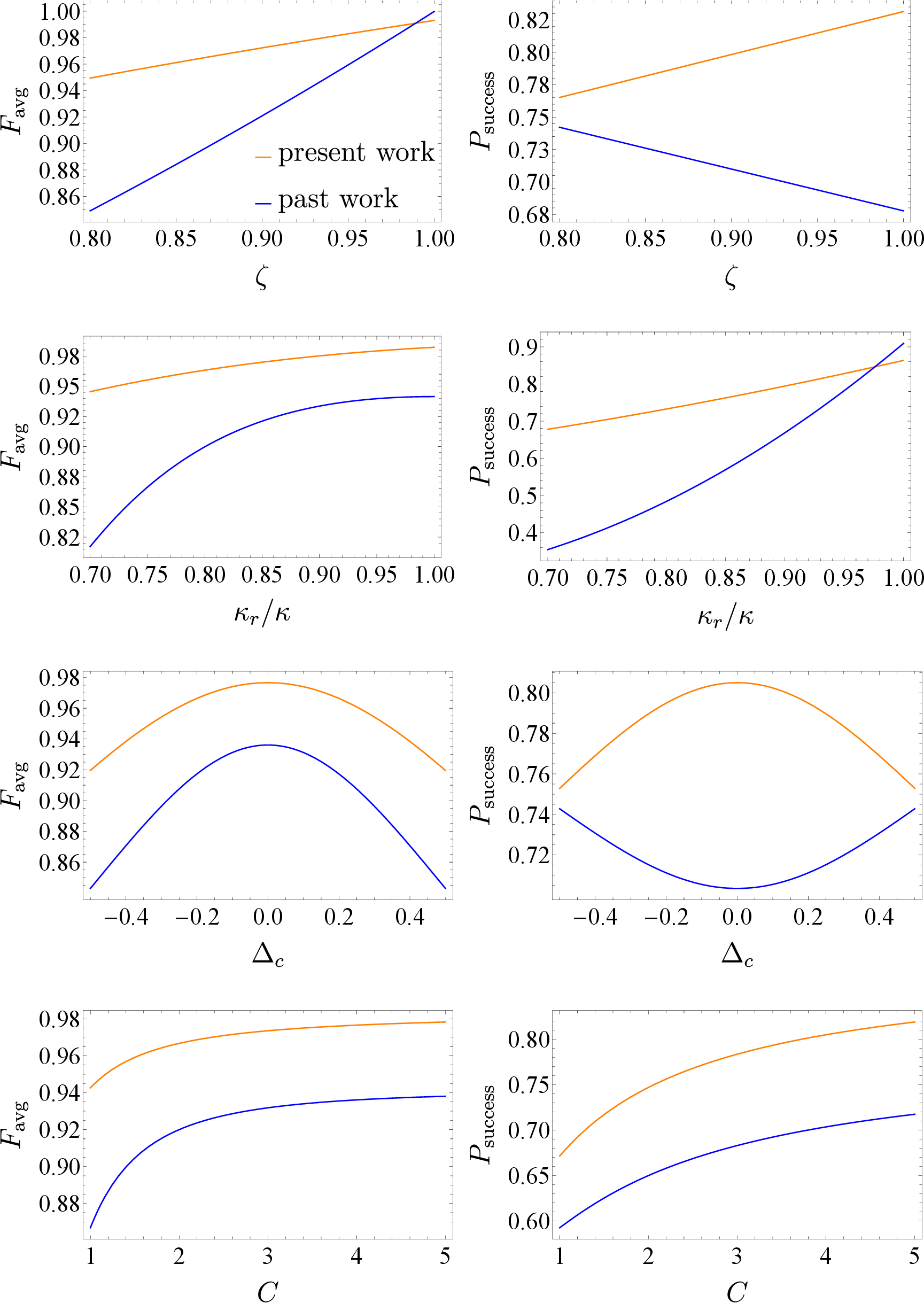}
    \caption{Average fidelity (\rsub{$F_{\rm avg}$}) and success probability ($P_{\rm success}$) for new (orange) and old (blue) atom-photon gate. $\zeta$ is the photon-cavity spatial matching efficiency, $\kappa_r$ the cavity decay rate into the desired mode, $\kappa$ the total cavity decay rate including losses, $\Delta_{\rm c}$ the fractional photon-cavity detuning, and $C$ the atom-cavity cooperativity. For each figure, all errors are fixed and one is varied. Values of errors used when they are fixed: $C=4$, $\kappa_r/\kappa=0.916$, and $\zeta=0.92$.}
    \label{CZerror}
\end{figure}

Previously, we have assumed no errors or losses in the atom-photon two-qubit gate operation. In this section we  estimate the reduction in the gate fidelity and success probability due to the following errors: spatial mode mismatch between the photon and the cavity, spontaneous emission, cavity losses, finite cooperativity, and photon-cavity/atom-cavity frequency detunings.

We will assume that the photon wavepacket has a narrow bandwidth around a central frequency $\omega_{\rm p}$ or equivalently that the temporal pulse width $T$ is large compared to $1/\kappa$, i.e., $\kappa T \gg 1$. Previous studies showed that this condition is required to ensure high fidelity and no photon pulse shape distortion \cite{Duan2004, Duan2005}.  

We analyze the $\sf CZ$ gate setup in Fig. \ref{CZ_gate}. We have an initial photon state $\alpha_{\rm p}\ket{V}+\beta_{\rm p}\ket{H}$ impinging on the cavity from the left on the atomic qubit $\alpha\ket{0}^{\rm a}+\beta\ket{1}^{\rm a}$. After the photon reflects off the cavity and exits through the bottom PBS, we end up with a $\sf CZ$ gate between the photon and the atom. Our scheme fails whenever the photon is lost from the cavity or when its polarization fails to flip during the scattering (and subsequently gets lost from the system). In the case that the photon is not lost, if we denote the final atom-photon state after the non-ideal gate operation as $\ket{\psi_{\text{out}}}$ and the output of an ideal gate as $\ket{\psi_{\text{ideal}}}$, then the fidelity is defined as $ F= |\braket{\psi_{\text{ideal}} | \psi_{\text{out}}}|^2$. \rsub{The success probability $ P_{\rm success}$ is the probability that the photon is not lost during the gate operation and available for detection. $P_{\rm success}=100 \%$ corresponds to a deterministic gate. Some applications require high $P_{\rm success}$, e.g., in a quantum network, it is crucial to have a large ratio between the remote entanglement success rate (which depends on $P_{\rm success}$) and the decoherence rate of the entangled qubits \cite{Covey2023}. Moreover, in quantum information processing applications, the efficiency of the scheme is an explicit function of $P_{\rm success}$ (e.g., proportional to $P_{\rm success}^2$ for remote entanglement). Thus, $F$ and $P_{\rm success}$ are measures of the gate quality and efficiency, respectively.}

The error analysis in Appendix \ref{CZ_error_new} provides analytic expressions for $F$ and $ P_{\rm success}$ that depend on the initial atom-photon amplitudes, the atom-cavity parameters, and the mode matching efficiency. As a reference, we compare our scheme to the one first proposed by Duan and Kimble \cite{Duan2004,Duan2005}, which has been subsequently experimentally implemented \cite{Reiserer2014,Welte2018, Daiss2021, Langenfeld2021}.  We introduce the same errors for their scheme and perform an analogous  error analysis in Appendix \ref{CZ_error_old}. To show the reliability of the model presented here, in Appendix \ref{predictpower} we analyze two previous atom-photon gate experiments using the error model from Appendix \ref{CZ_error_old}. We show that the error model has good predictive power in estimating reductions in fidelity and success probability in actual experiments (due to the effects considered), where the agreement between our error model and the experiments is within one to two percent.

In Fig. \ref{CZerror}, we show the results of the error analysis for the atom-photon $\sf CZ$ gate for the present (orange) and previous (blue) schemes. The figures on the left are the average fidelity \rsub{$F_{\rm avg}$} and on the right are the average success probability $P_{\rm success}$, averaged over all possible initial atom-photon product states. We plot \rsub{$F_{\rm avg}$} and $P_{\rm success}$ versus four experimental parameters of interest: $\zeta$ the photon-cavity spatial mode matching efficiency, $\kappa_r/\kappa$, which captures cavity mirror losses ($\kappa_r$ is the cavity decay rate into the desired mode and $\kappa$ is the total cavity decay rate including losses), $\Delta_{\rm c}$ the fractional photon-cavity detuning, and $C$ the atom-cavity cooperativity. The values and ranges chosen for $(\zeta,\kappa_r/\kappa,\Delta_{\rm c},C)$ are from previous atom-photon gate experiments (see also Appendix \ref{predictpower} for details) \cite{Welte2018, Reiserer2014}. 

First, consider spatial mode mismatch, which is the largest source of infidelity in atom-photon gates \cite{Welte2018, Reiserer2014}. As the spatial mode matching efficiency $\zeta$ decreases from $100\%$ to $80\%$, the fidelity in the older scheme decreases by $15\%$ while it only decreases by $4\%$ in the present scheme (see Fig. \ref{CZerror}). Another parameter of interest is cavity losses $\kappa_r/\kappa$. As $\kappa_r/\kappa$ decreases from $1$ to $0.7$, the fidelity in the older scheme decreases by $12.4\%$ while it only decreases by $3.7\%$ in the present scheme.  The error analysis shows that the fidelity and success probability have similar behavior versus $C$ and $\Delta_{\rm c}$ both in the older and the present scheme. All in all, Fig. \ref{CZerror} shows that, in general, the present scheme has higher average fidelity and success probability than the previous scheme. Moreover, the new scheme is more robust against experimental errors like losses, mismatch, and detunings. \rsub{In Appendix \ref{loss_fidelity}, we elaborate on the connection between  the $\sf CZ$ gate fidelity and photon loss in the present scheme and propose how to increase the fidelity further}. 

It is worth explaining why $P_{\rm success}$ in the present and the previous schemes have opposite behavior versus $\zeta$ and $\Delta_{\rm c}$. First, consider the behavior versus $\zeta$. If the mode matching efficiency is $\zeta$, it is empirically found that the mismatched part of the optical mode  $1-\zeta$ reflects completely without any change or interaction with the cavity, and thus does not lead to a $\sf Z$ gate. Under the present scheme, failure to perform the $\sf Z$ gate is rejected from the system, which translates to loss (see Sec. \ref{Atom-photon two-qubit gate}). Therefore, the more there is a mismatch (i.e., lower $\zeta$), the lower the success probability. Under the previous scheme, the mismatched part is not rejected from the system so it does not translate into loss. If there is more mismatch, then there is a smaller fraction of the total light that interacts with the cavity (i.e., the matched part, which can get lost through spontaneous emission, scattering, or transmission) and a larger fraction that gets completely reflected back (i.e., mismatched part), which translates into higher success probability (albeit lower fidelity). The behavior versus $\Delta_{\rm c}$ is explained in a similar fashion. For sufficiently large $\Delta_{\rm c}$, the photon is essentially not coupled to the atom-cavity system, regardless of the atomic qubit state. Under the present scheme, this leads to a failure in the $\sf Z$ gate, which again is rejected and translates into loss. Under the previous scheme, failure to perform the $\sf Z$ gate is not rejected from the system. Therefore, a photon that is not coupled will be less likely to get lost through spontaneous emission or scattering off the atom, which increases the success probability.

\begin{figure}[!t] 
    \centering
    \includegraphics[width=0.5\textwidth]{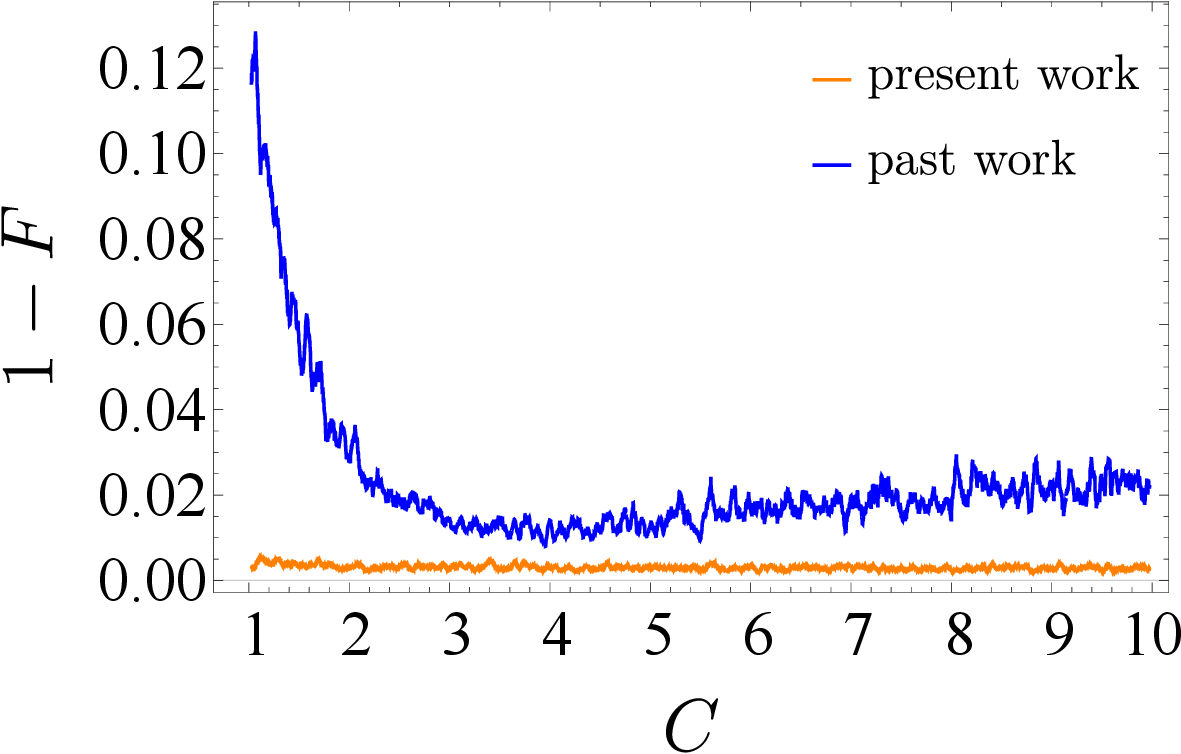}
    \caption{The infidelity $1-F$ in atom-atom entanglement in the present (orange) and previous (blue) work versus $C$ (assuming no spatial mode mismatch). The following parameters for the Gaussian distributions were used ($\overline{X}$ denotes the average of $X$ and $\sigma_X$ denotes the standard deviation): $\overline{\kappa_r/\kappa}=0.9,  \sigma_{\kappa_r/\kappa}=0.05, \overline{\Delta_{\rm c}}= \overline{\Delta_{\rm a}}=0,$ and $\sigma_{\Delta_{\rm c}}=\sigma_{\Delta_{\rm a}}=0.05$. As we vary $C$ in this plot, we choose $\overline{C}=C$, and $\sigma_{\rm c}=0.1 C$. This figure is a moving average of each 50 neighboring points $(C,1-F)$.}
    \label{Fatom}
\end{figure}

Next, we investigate the infidelity $1-F$ of remote atom-atom entanglement due to random cavity parameter variations and frequency fluctuations. The setup for generating atom-atom entanglement in our scheme is shown in Fig. \ref{atom_atom}. In Appendices \ref{CZ_error_new} and \ref{CZ_error_old}, we derive the entanglement fidelity as a function of the parameters of the two cavities ($C, \Delta_{\rm c}, \Delta_{\rm a},$ and $\kappa_r/\kappa$), assuming no spatial mode mismatch. We characterize the robustness of the resultant atom-atom entanglement against cavity parameter variations and frequency fluctuations as follows. We treat all the cavity parameters and the frequency fluctuations as random Gaussian variables, and compute the infidelity both in our scheme and the previous scheme. The parameters of the Gaussian distribution (average and standard deviation) are chosen such that they are close to the values of recent experiments \cite{Reiserer2014, Daiss2021,Welte2018, Welte2021}. The result is shown in Fig. \ref{Fatom}, where we plot the average infidelity $1-F$ versus $C$ for the present (orange) and previous scheme (blue), and we randomly vary all the parameters of the two cavities (assuming Gaussian distributions). In the previous scheme, the average infidelity is large for small $C$ ($C<3$), and for higher $C$ it fluctuates around $2\%$. The present scheme is less sensitive to the various random fluctuations, where the infidelity varies around $0.3\%$ regardless of $C$.

\begin{figure}[!t] 
    \centering
    \includegraphics[width=0.5\textwidth]{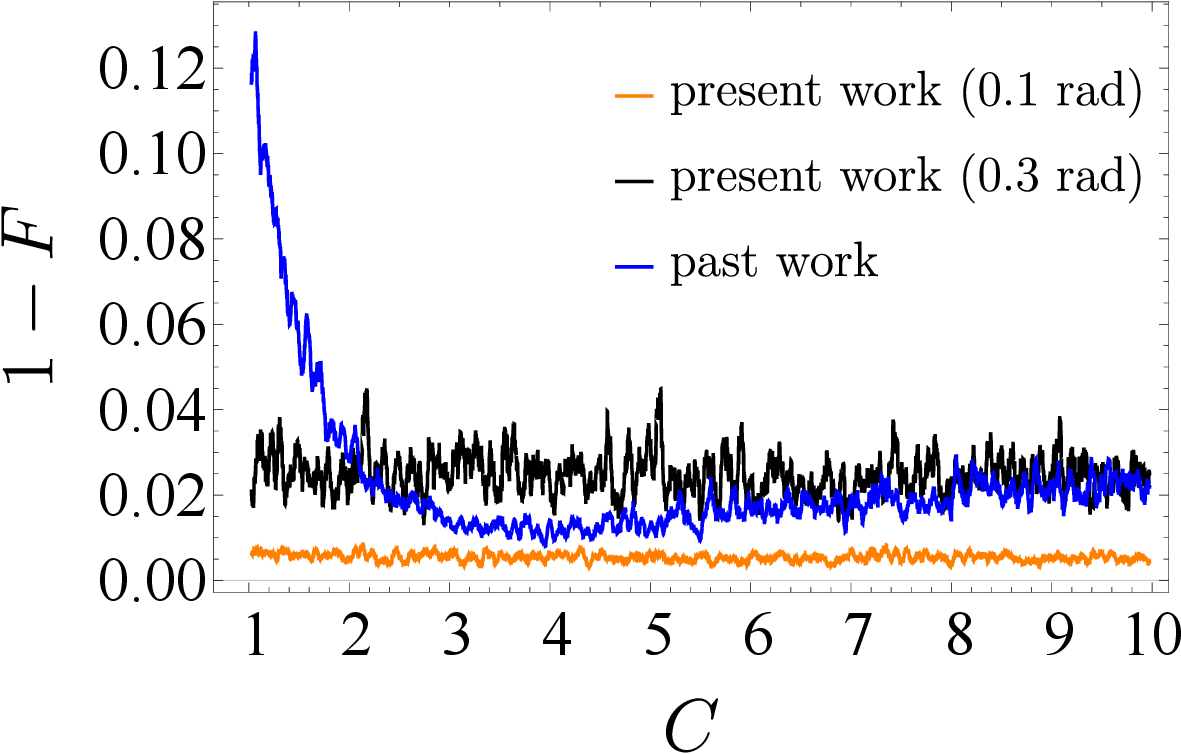}
    \caption{Same as Fig. \ref{Fatom} but with the addition of phase fluctuation in the present scheme (orange and black). The phase fluctuation $\phi$ introduced is a Gaussian random variable with $\overline{\phi}=0$ with   $\sigma_{\phi}=0.1$ radians (orange) and $\sigma_{\phi}=0.3$ radians (black).  }
    \label{Fatom2}
\end{figure}

In Fig. \ref{Fatom2}, in addition to the cavity parameters and frequency fluctuations introduced earlier, we introduce random phase fluctuation in the arms of the MZI in the atom-photon gate (see Figs. \ref{atom_atom} and \ref{CZ_gate}). This phase fluctuation acts as a dephasing error and increases the infidelity in our scheme. For small phase fluctuations (below $0.1$ radians), $1-F$ remains below $1\%$ in the present scheme (orange). However, for sufficiently large phase fluctuations (above $0.3$ radians), the present scheme (black) performs comparably or worse than the previous scheme (blue) for large $C$. This shows the importance of interferometric stability in the present scheme to achieve high fidelity.

\section{Discussion}\label{Discussion}

Our analysis assumed a single photon source, perfect photon detection, and ignored other error effects. These effects will further reduce both the fidelity and the success probability in both schemes. Multiphoton effects stemming from the use of a weak coherent photon source are particularly important. In recent experiments, typically a weak coherent source of photons with a Poissonian distribution is used, i.e., $P(n)=\bar{n}^n e^{-\bar{n}}/n!$, where $\bar{n}$ is the mean photon number and $P(n)$ is the probability to have $n$ photons. For example, consider an experiment to generate atom-photon Bell states (starting from a photon and an atom both in an equal superposition state) with the parameters \cite{Welte2018}: $\bar{n}=0.13$, an average single photon detection efficiency of $\eta=55\%$, $\zeta=0.92$, $C=4$, and $\kappa_r/\kappa= 0.916$. The total success probability to generate atom-photon Bell states is then reduced from $P_{\rm success}= 70\% (80\%)$ in the previous (new)  schemes, respectively to $P(1)\eta P_{\rm success}=4.4\%(5\%)$,  in the previous (new)  schemes, respectively. Thus, we expect that the success probability will be comparable in both schemes. To significantly improve the success probability, using single photon sources and better single photon detectors is necessary.

\rsub{
There are two categories of errors that contribute to the infidelity of the atom-photon gate. The first is scheme dependent errors related to the gate operation itself (e.g., the errors considered in this work). The second is not directly related to the gate operation but still causes infidelities (e.g., non-ideal single photon source, non-ideal photon detection, multi-photon effects, atomic state preparation errors, readout, detector dark counts, atomic decoherence, and non-ideal optical components). The relative contribution of these error categories depend on the experiment itself \cite{Daiss2021,Welte2018,Langenfeld2021,Welte2021,Welte2018, Reiserer2013,Langenfeld2021, Hacker2019}, but both generally contribute significantly to the infidelity (e.g., in one atom-photon gate experiment \cite{Reiserer2014}, the infidelity contributions of the two error types were $8(3)\%$ and $10\%$, respectively, see also Appendix \ref{predictpower}). The contribution of the present work is a proposal to reduce the first error type, i.e., a higher atom-photon gate quality. To achieve very high fidelities ($\ge 99 \%$), significant effort in experimentally eliminating the second error category is also necessary.
}

In summary, we introduced a new scheme to perform atom-photon gates using symmetric energy levels. Error analysis shows that the gate fidelity is more robust against errors like spatial mode mismatch between the photon and the cavity, spontaneous emission, cavity losses, finite cooperativity, photon-cavity and atom-cavity frequency detunings. Moreover, the fidelity of the atom-photon gate does not depend on the state of the  atomic qubit, a desirable feature in quantum information processing applications. The gate robustness is advantageous in other contexts such as  remote atom-atom entanglement, where the fidelity is less sensitive to random variations of the parameters of the cavities and frequency fluctuations. A potential bottleneck in this scheme is interferometric stability, where phase stabilization is required to ensure high fidelity gate operations. The error analysis supports the feasibility of reaching high fidelity with only modest requirements on the interferometric phase stability, that are compatible  with  current technology.

\section*{Acknowledgments}
We would like to thank Anirudh Yadav for valuable discussions. This material is based upon work supported by the
U.S. Department of Energy Office of Science National
Quantum Information Science Research Centers as part of the Q-NEXT center, as well as support from NSF Award 2016136 for the QLCI center Hybrid
Quantum Architectures and Networks, and NSF award 2228725. 


\bibliography{rydberg,optics,qc_refs,atomic,saffman_refs}

\begin{widetext}

\appendix

\section{Implementation with $^{87}\rm Rb$}\label{Rb_implement}

The atomic energy level scheme in Fig. 1c)  can be implemented with $^{87}\rm Rb$ using $\ket{F=2,m_F=\pm2}$ and $\ket{F'=3,m_{F'}=\pm3}$ for the lower and upper states, respectively, i.e., $\ket{g_{\pm}}$ are $5\, ^{2}S_{1/2}\ket{2,\pm2}$   and $\ket{e_{\pm}}$ are $5  ^{2}P_{3/2}\ket{3,\pm3}$. The states $\ket{F'=3,|m_{F'}|<3}$ can be shifted away from the $F=2 \leftrightarrow F'=3$ resonance to isolate the stretched state transitions by means of the tensor shift imparted by a dipole trap with linear polarization parallel or perpendicular to the quantization axis \cite{Reiserer2014}. In particular, a 1064 dipole trap polarized parallel to the quantization axis with a depth of 2.5 mK   will induce shifts relative to $\ket{F=3,m_F=\pm3}$ of 52 MHz for $\ket{F=3,m_F=\pm2}$, 83 MHz for $\ket{F=3,m_F=\pm1}$, and 93 MHz for $\ket{F=3,m_F=\pm0}$ \cite{LeKien2013}. These shifts are sufficient for ensuring negligible coupling to states in $F=3$ with $|m_F|<3$. The Land\'e $g_F$ factors for the $F=2$ and $F'=3$ levels are 0.7 MHz/G and 0.93 MHz/G, respectively. 
This gives a $\pm$1.39 MHz/G first order differential Zeeman shift for the two transitions, or a differential sensitivity of 2.78 MHz/G.

Typical magnetic noise is on the order of 1 mG. Magnetic stability better than 50 $\rm \mu$G has been demonstrated using an active feedforward approach \cite{Merkel2019}. Magnetic noise can also affect the coherence of the atomic qubit, which is encoded in the magnetically sensitive $\ket{F=2,m_F=\pm2}$ states. This sensitivity can be reduced using microwave or radio frequency dressing \cite{Sarkany2014,Sinuco-Leon2021}. Operating with a bias field of 0.5 G and 1 mG rms noise, by applying a single microwave dressing field linearly polarized along $x$, with detuning $\Delta/2\pi = 1$ MHz from the hyperfine splitting frequency (6.834 GHz) and Rabi frequency $\Omega/2\pi$ = 23.47 kHz, $T_2^*$ can exceed 8 ms, compared with 0.5 ms with no dressing for the same bias field and noise. Finally, the equal superposition state $\ket{0}_{x}^{\rm a}=(\ket{0}^{\rm a}+\ket{1}^{\rm a})/\sqrt{2}$ can be generated by a sequence of $\pi$ and $\pi/2$ pulses \cite{Thomas2022}.

\section{Fidelity and success probability analysis}\label{error}

\subsection{Present work} \label{CZ_error_new}

\subsubsection{Atom-photon $\sf CZ$ gate}

The setup is shown in Fig. \ref{CZ_gate}. We start with the photon-atom product state:

\begin{eqnarray}  
\nonumber \ket{\psi}&=&(\alpha_{\rm p}\ket{V}+\beta_{\rm p}\ket{H})(\alpha\ket{0}+\beta\ket{1}) \\
&=&\alpha_{\rm p}\ket{V}(\alpha\ket{0}^{\rm a}+\beta\ket{1}^{\rm a})+\beta_{\rm p}\ket{H}(\alpha\ket{0}^{\rm a}+\beta\ket{1}^{\rm a})
\end{eqnarray}

The first PBS separates $\ket{H}$ and $\ket{V}$ into paths $r_1$ and $p_2$ respectively:

\begin{align}  
 \ket{\psi}=\alpha_{\rm p}\ket{V}_{p_2}(\alpha\ket{0}^{\rm a}+\beta\ket{1}^{\rm a})+\beta_{\rm p}\ket{H}_{r_1}(\alpha\ket{0}^{\rm a}+\beta\ket{1}^{\rm a})
\end{align}

The photon in path $r_1$ scatters off the cavity under the following scattering relations:

\begin{eqnarray} \label{scatter_relations1}
\ket{0}^{\rm a}\ket{\sigma_+} &\rightarrow& r_{\rm c}\ket{0}^{\rm a}\ket{\sigma_+}+t_{\rm c} \ket{L} \\ \label{scatter_relations2}
\ket{1}^{\rm a}\ket{\sigma_-} &\rightarrow& r_{\rm c}\ket{1}^{\rm a}\ket{\sigma_-}+t_{\rm c} \ket{L'} \\ \label{scatter_relations3}
\ket{0}^{\rm a}\ket{\sigma_-} &\rightarrow&  r_{\rm nc}\ket{0}^{\rm a}\ket{\sigma_-}+t_{\rm nc} \ket{L''}\\
\ket{1}^{\rm a}\ket{\sigma_+} &\rightarrow& r_{\rm nc}\ket{1}^{\rm a}\ket{\sigma_+}+t_{\rm nc} \ket{L'''} \label{scatter_relations4}
\end{eqnarray}

where \cite{Reiserer2015}

\begin{eqnarray} 
r_{\rm c}&=&1-\dfrac{2\dfrac{\kappa_r}{\kappa}(i\Delta_{\rm a}+1)}{(i\Delta_{\rm c}+1)(i\Delta_{\rm a}+1)+2C}\\
r_{\rm nc}&=&1-\dfrac{2\dfrac{\kappa_r}{\kappa}}{i\Delta_{\rm c}+1}
\end{eqnarray}

Here we take into account that not all incident light is reflected from the cavity ($|r_{\rm c}|,|r_{\rm nc}|<1$), because there is a probability to lose the photon due to spontaneous emission into free space, transmission through the highly reflective mirror, scattering with the atom or the cavity mirrors into free space. Note that $\kappa=\kappa_r+\kappa_{\text{loss}}$, where $\kappa_r$ is the cavity decay rate through the coupling mirror (desired mode), while $\kappa_{\text{loss}}$ encompasses the contributions due to loss by transmission through the other mirror or scattering off the mirrors. To take losses after scattering into account, we introduce the ``loss" states $\ket{L},\ket{L'},\ket{L''},$ and $\ket{L'''}$ with amplitudes $t_{\rm c}$ and $t_{\rm nc}$. By probability conservation, the probabilities to lose the photon are then $|t_{\rm c}|^2=1-|r_{\rm c}|^2$ and $|t_{\rm nc}|^2=1-|r_{\rm nc}|^2$ for the coupled and uncoupled cases.

We would like to add another important source of nonideality, namely spatial mode mismatch between the photon and the cavity mode. If the mode matching efficiency is $\zeta$, it is empirically found that the mismatched part reflects completely without any change. Thus we need to modify our reflection coefficients as \cite{Reiserer2015}

\begin{equation}
|r|^2 \rightarrow (1-\zeta)+\zeta|r|^2
\end{equation}

To take this into account, we introduce  two orthogonal spatial modes $\ket{\text{mat}}$ (``matched") and $\ket{\text{mis}}$ (``mismatched") with the following relative amplitudes:

\begin{equation}\label{mismatch}
\sqrt{1-\zeta} e^{i\theta}\ket{\text{mis}}+\sqrt{\zeta}\ket{\text{mat}}
\end{equation}

Thus, we divide the entire initial quantum state into matched and mismatched modes:

\begin{equation}
 \ket{\psi}=\ket{\phi}_{p_2}+\sqrt{1-\zeta}e^{i\theta}\beta_{\rm p}\ket{H,\text{mis}}_{r_1}(\alpha\ket{0}^{\rm a}+\beta\ket{1}^{\rm a})+\sqrt{\zeta}\beta_{\rm p}\ket{H,\text{mat}}_{r_1}(\alpha\ket{0}^{\rm a}+\beta\ket{1}^{\rm a})
\end{equation}
where
\begin{equation}
 \ket{\phi}_{p_2}=\sqrt{1-\zeta}e^{i\theta}\alpha_{\rm p}\ket{V,\text{mis}}_{p_2}(\alpha\ket{0}^{\rm a}+\beta\ket{1}^{\rm a})+\sqrt{\zeta}\alpha_{\rm p}\ket{V,\text{mat}}_{p_2}(\alpha\ket{0}^{\rm a}+\beta\ket{1}^{\rm a})
\end{equation}

Expanding the matched part $\ket{H,\text{mat}}_{r_1}$ in terms of $\ket{\sigma_{\pm},\text{mat}}_{r_1}$:

\begin{eqnarray}
 \ket{\psi}&=&\ket{\phi}_{p_2}+\sqrt{1-\zeta}e^{i\theta}\beta_{\rm p}\ket{H,\text{mis}}_{r_1}(\alpha\ket{0}^{\rm a}+\beta\ket{1}^{\rm a})\nonumber\\
 &+&\sqrt{\zeta}\dfrac{\beta_{\rm p}}{\sqrt{2}}\bigg\{ \ket{\sigma_+,\text{mat}}_{r_1}(\alpha \ket{0}^{\rm a}+\beta \ket{1}^{\rm a}) +\ket{\sigma_-,\text{mat}}_{r_1}(\alpha \ket{0}^{\rm a}+\beta \ket{1}^{\rm a})\bigg\}
\end{eqnarray}

Only the matched mode in $r_1$ experiences the scattering relations Eqs.  (\ref{scatter_relations1} -\ref{scatter_relations4}). The state after scattering then becomes: 

\begin{multline}
 \ket{\psi}=\ket{\phi}_{p_2}+\sqrt{1-\zeta}e^{i\theta}\beta_{\rm p}\ket{H,\text{mis}}_{r_1}(\alpha\ket{0}^{\rm a}+\beta\ket{1}^{\rm a})\\+\sqrt{\zeta}\dfrac{\beta_{\rm p}}{\sqrt{2}}\bigg\{ \ket{\sigma_+,\text{mat}}_{r_1}(\alpha r_{\rm c}\ket{0}^{\rm a}+\beta r_{\rm nc} \ket{1}^{\rm a}) +\ket{\sigma_-,\text{mat}}_{r_1}(\alpha r_{\rm nc}\ket{0}^{\rm a}+\beta r_{\rm c}\ket{1}^{\rm a})+\alpha t_{\rm c}\ket{L}+\beta t_{\rm c}\ket{L'}+\alpha t_{\rm nc}\ket{L''}+\beta t_{\rm nc}\ket{L'''}\bigg\}
\end{multline}

The total probability to lose the photon out of the cavity into free space is then

\begin{equation} 
 P_{\text{loss}}=P(\ket{L})+P(\ket{L'})+P(\ket{L''})+P(\ket{L'''})=\zeta \dfrac{|\beta_{\rm p}|^2}{2}(|t_{\rm c}|^2+|t_{\rm nc}|^2)=\zeta \dfrac{|\beta_{\rm p}|^2}{2}(2-|r_{\rm c}|^2-|r_{\rm  nc}|^2)
\end{equation}

where $|t|^2=1-|r|^2$ and we assumed that the lost states are orthogonal (the validity of this assumption and our error model more generally is evaluated in Appendix \ref{predictpower}). Since we are interested in the case where the photon is detected and the scheme succeeds, we consider the quantum state when it gets projected into the ``not lost" state and gets reflected out of the cavity:

\begin{eqnarray}
 \ket{\psi}&=&\dfrac{1}{N_l}\bigg[\ket{\phi}_{p_2}+\sqrt{1-\zeta}e^{i\theta}\beta_{\rm p}\ket{H,\text{mis}}_{r_1}(\alpha\ket{0}^{\rm a}+\beta\ket{1}^{\rm a})\nonumber\\
 &+&\sqrt{\zeta}\dfrac{\beta_{\rm p}}{\sqrt{2}}\bigg\{ \ket{\sigma_+,\text{mat}}_{r_1}(\alpha r_{\rm c}\ket{0}^{\rm a}+\beta r_{\rm  nc} \ket{1}^{\rm a}) +\ket{\sigma_-,\text{mat}}_{r_1}(\alpha r_{\rm  nc}\ket{0}^{\rm a}+\beta r_{\rm c}\ket{1}^{\rm a})\bigg\}\bigg]
\end{eqnarray}

where $N_l=\sqrt{1-P_{\rm loss}}$ is the normalization constant. We can rewrite the state as:

\begin{eqnarray} 
 \ket{\psi}&=&\dfrac{1}{N_l}\bigg[\ket{\phi}_{p_2}+\sqrt{1-\zeta}e^{i\theta}\beta_{\rm p}\ket{H,\text{mis}}_{r_1}(\alpha\ket{0}^{\rm a}+\beta\ket{1}^{\rm a})\nonumber\\
 &+&\sqrt{\zeta}\beta_{\rm p}\bigg\{\dfrac{r_{\rm c}-r_{\rm  nc}}{2} \ket{V,\text{mat}}_{r_1}(\alpha \ket{0}^{\rm a}-\beta \ket{1}^{\rm a})+\dfrac{r_{\rm c}+r_{\rm  nc}}{2}\ket{H,\text{mat}}_{r_1} (\alpha \ket{0}^{\rm a}+\beta\ket{1}^{\rm a})\bigg\}\bigg]
\end{eqnarray}

There is another source of photon loss. After reflection, there is a probability that the photon polarization does not flip during the scattering and remains in $\ket{H}_{r_1}$. This component will not be reflected down by the PBS to the path $d_1$, and it will be lost from the system. The probability for this to happen is  

\begin{equation} 
 P(H)=\dfrac{1}{|N_l|^2}\bigg[(1-\zeta)|\beta_{\rm p}|^2+\zeta \dfrac{|\beta_{\rm p}|^2}{4}|r_{\rm c}+r_{\rm  nc}|^2\bigg]
\end{equation}

In the event that the photon is detected, the quantum state will be projected to the state without $\ket{H}_{r_1}$ component:

\begin{equation} 
 \ket{\psi}=\dfrac{1}{N_l N_h}\bigg[\ket{\phi}_{p_2}+\sqrt{\zeta}\beta_{\rm p}\dfrac{r_{\rm c}-r_{\rm  nc}}{2} \ket{V,\text{mat}}_{r_1}(\alpha \ket{0}^{\rm a}-\beta \ket{1}^{\rm a})\bigg]
\end{equation}

where $N_h=\sqrt{1-P(H)}$ is the normalization constant. After scattering, $\ket{V}_{r_1}$ is reflected down to path $d_1$ by the PBS. Moreover, the HWP in $d_1$ flips the polarization of the photon to $\ket{H}_{d_1}$. Thus we have

\begin{equation} 
 \ket{\psi}=\dfrac{1}{N_l N_h}\bigg[\ket{\phi}_{p_2}+\sqrt{\zeta}\beta_{\rm p}\dfrac{r_{\rm c}-r_{\rm  nc}}{2} \ket{H,\text{mat}}_{d_1}(\alpha \ket{0}^{\rm a}-\beta \ket{1}^{\rm a})\bigg]
\end{equation}

Finally, after we combine the two paths of the photon by the PBS, we get the output state:

\begin{equation} 
 \ket{\psi}=\dfrac{1}{N_l N_h}\bigg[\ket{\phi}+\sqrt{\zeta}\beta_{\rm p}\dfrac{r_{\rm c}-r_{\rm  nc}}{2} \ket{H,\text{mat}}(\alpha \ket{0}^{\rm a}-\beta \ket{1}^{\rm a})\bigg]
\end{equation}

where we have removed the path information ($d_1$ and $p_2$). Compare this with the output of an ideal $\sf CZ$:

\begin{equation} 
 \ket{\psi_{\text{ideal}}}=\alpha_{\rm p}\ket{V}(\alpha\ket{0}^{\rm a}+\beta\ket{1}^{\rm a})+\beta_{\rm p} \ket{H}(\alpha \ket{0}^{\rm a}-\beta \ket{1}^{\rm a})
\end{equation}

Thus our output state is

\begin{equation} 
 \ket{\psi_{\text{out}}}=\dfrac{1}{N_l N_h}\bigg[\sqrt{1-\zeta}e^{i\theta}\alpha_{\rm p}\ket{V,\text{mis}}(\alpha\ket{0}^{\rm a}+\beta\ket{1}^{\rm a})+ \sqrt{\zeta}\ket{\psi_{\text{ideal}},\text{mat}}+\sqrt{\zeta}\ \dfrac{\beta_{\rm p}}{2}\ket{H,\text{mat}} (r_{\rm c}-r_{\rm nc}-2)(\alpha \ket{0}^{\rm a}-\beta\ket{1}^{\rm a})\bigg]
\end{equation}

The fidelity is defined as 
$ F= |\braket{\psi_{\text{ideal}} | \psi_{\text{out}}}|^2$
where after taking the inner product $\braket{\psi_{\text{ideal}} | \psi_{\text{out}}}$, we trace over both the matched and mismatched modes (i.e., when the photon gets detected it is projected to a state of definite polarization and spatial mode). Using all the relevant equations above, we get:

\begin{equation} 
 F= \dfrac{1}{|N_l N_h|^2}\bigg\{ (1-\zeta ) |\alpha_{\rm p}|^4+\zeta \bigg|1+|\beta_{\rm p}|^2\dfrac{(r_{\rm c}-r_{\rm nc}-2)}{2}\bigg|^2\bigg\}
\end{equation}

If there is a phase difference $\phi$ between the arms of the MZI, then the fidelity gets modified to: 

\begin{equation} 
 F= \dfrac{1}{|N_l N_h|^2}\bigg\{ (1-\zeta ) |\alpha_{\rm p}|^4+\zeta \bigg||\alpha_{\rm p}|^2e^{i\phi}+|\beta_{\rm p}|^2+|\beta_{\rm p}|^2\dfrac{(r_{\rm c}-r_{\rm nc}-2)}{2}\bigg|^2\bigg\}
\end{equation}

No photon is detected if it is lost ($P_{\rm loss}$) or the polarization fails to flip and remains horizontal after scattering [$P(H)$]. Therefore, our success probability is

\begin{equation} 
 P_{\rm success}= 1-P_{\rm fail}=1-[P_{\rm loss}+(1-P_{\rm loss})P(H)]=|N_l N_h|^2
\end{equation}

\rsub{
To get the average fidelity $F_{\rm avg}$, $F$ is averaged over all possible initial atom-photon product states. Noting that here $F$  only depends on the photon's initial amplitude, the average is taken over the Bloch sphere of the photon by making the substitution $\alpha_{\rm p} \rightarrow \cos (\theta/2) e^{i \Phi}$ and $\beta_{\rm p} \rightarrow \sin (\theta/2)$ in $F$:}

\begin{equation} 
\rsub{ F_{\rm avg}= \dfrac{1}{4 \pi} \int_{0}^{\pi} \int_0^{2 \pi} d\theta d\Phi\,  F \sin \theta }.
\end{equation}

\subsubsection{Remote atom-atom entanglement}

Here we start with the photon-atom-atom product state $\ket{\sigma_+}\ket{0}^{\rm a}_x \ket{1}^{\rm a}_x$ (see Fig. \ref{atom_atom}). For simplicity, we assume no spatial mode mismatch. After the first $\sf CZ$ gate between $\ket{\sigma_+}$ and $\ket{0}^{\rm a}_x$, we get (ignoring normalization and assuming the photon was not lost):

\begin{equation} 
\big[\ket{V}\ket{0}^{\rm a}_x+\dfrac{r_{\rm c}-r_{\rm nc}}{2}\ket{H}\ket{1}^{\rm a}_x \big]\ket{1}^{\rm a}_x
\end{equation}

After applying the HWP, we have

\begin{equation} 
\big(\ket{H}\ket{0}^{\rm a}_x+\dfrac{r_{\rm c}-r_{\rm nc}}{2}\ket{V}\ket{1}^{\rm a}_x \big)\ket{1}^{\rm a}_x
\end{equation}

We send the photon to the second cavity on the right. After the second $\sf CZ$ gate, we get:

\begin{equation} 
(r'_{\rm c}-r'_{\rm nc})\ket{H}\ket{00}_x^{\rm a} +(r_{\rm c}-r_{\rm nc})\ket{V}\ket{11}_x^{\rm a} 
\end{equation}

where $r'_{\rm c}-r'_{\rm nc}$ are the reflection coefficients of the second cavity, which in general are different from that of the first cavity $r_{\rm c}-r_{\rm nc}$. After the photon passes through a QWP, we get:

\begin{equation} 
\ket{V}\big \{(r'_{\rm c}-r'_{\rm nc})\ket{00}_x^{\rm a} +(r_{\rm c}-r_{\rm nc})\ket{11}_x^{\rm a}\big  \}-\ket{H}\big  \{(r'_{\rm c}-r'_{\rm nc})\ket{00}_x^{\rm a} -(r_{\rm c}-r_{\rm nc})\ket{11}_x^{\rm a} \big  \}
\end{equation}

Measuring the photon polarization results in the following atom-atom entangled states:

\begin{eqnarray} 
\ket{V}&:& (r'_{\rm c}-r'_{\rm nc})\ket{00}_x +(r_{\rm c}-r_{\rm nc})\ket{11}_x\\
\ket{H}&:& (r'_{\rm c}-r'_{\rm nc})\ket{00}_x^{\rm a} -(r_{\rm c}-r_{\rm nc})\ket{11}_x^{\rm a}
\end{eqnarray}

If the two cavities are identical, i.e., $r'_{\rm c}-r'_{\rm nc}=r_{\rm c}-r_{\rm nc}$, then we get the maximally entangled Bell states:

\begin{eqnarray} 
\ket{V}&:& \ket{00}_x^{\rm a} +\ket{11}_x^{\rm a}
\\
\ket{H}&:& \ket{00}_x^{\rm a} -\ket{11}_x^{\rm a}
\end{eqnarray}

These Bell states correspond to $\ket{00}^{\rm a} +\ket{11}^{\rm a}$ and $\ket{01}^{\rm a} +\ket{10}^{\rm a}$ in the $z$ basis, respectively. If the two cavities are not identical, then the fidelity of the atom-atom Bell states is:

\begin{equation} 
F=\dfrac{1}{2}\dfrac{|(r'_{\rm c}-r'_{\rm nc})+(r_{\rm c}-r_{\rm nc})|^2}{|r'_{\rm c}-r'_{\rm nc}|^2+|r_{\rm c}-r_{\rm  nc}|^2}
\end{equation}

If there is a phase difference $\phi_1$ and $\phi_2$ between the arms of the first and second MZI that implement the $\sf CZ$ gates, then the fidelity will be modified as:

\begin{equation} 
F=\dfrac{1}{2}\dfrac{|(r'_{\rm c}-r'_{\rm  nc})+(r_{\rm c}-r_{\rm  nc})e^{i(\phi_2-\phi_1)}|^2}{|r'_{\rm c}-r'_{\rm  nc}|^2+|r_{\rm c}-r_{\rm  nc}|^2}
\end{equation}

\subsection{Previous schemes}\label{CZ_error_old}

\subsubsection{Atom-photon $\sf CZ$ gate}

We perform a similar error analysis for the scheme first proposed by Duan and Kimble \cite{Duan2004, Duan2005}. Subsequent schemes are similar or slightly modified variations \cite{An2009, QChen2009, GYWang2016}, and it have also been experimentally implemented \cite{Reiserer2014,Welte2018, Daiss2021}. \rsub{More precisely, we analyze a variant of the Duan and Kimble scheme that was implemented by Reiserer et al. \cite{Reiserer2014} as there is available experimental data for it. This allows us to evaluate the predictive power of our error model in Appendix \ref{predictpower}}. This scheme for atom-photon gates uses three energy levels: two ground states $\ket{0}^{\rm a}$ and $\ket{1}^{\rm a}$ that are not degenerate, and an excited state $\ket{e}^{\rm a}$. $\ket{1}^{\rm a}$ couples to $\ket{e}^{\rm a}$ by $\ket{\sigma_+}$ while $\ket{0}^{\rm a}$ is far-detuned. Here the photon path is not separated by an MZI, but the entire photon hits and reflects off the cavity. This gives the following scattering relations between the photon and the atom:

\begin{eqnarray} 
\ket{0}^{\rm a}\ket{\sigma_+} &\rightarrow& r_{\rm  nc}\ket{0}^{\rm a}\ket{\sigma_+}+t_{\rm  nc} \ket{L} \\
\ket{1}^{\rm a}\ket{\sigma_-} &\rightarrow& r_{\rm  nc}\ket{1}^{\rm a}\ket{\sigma_-}+t_{\rm  nc} \ket{L'} \\
\ket{0}^{\rm a}\ket{\sigma_-} &\rightarrow&  r_{\rm  nc}\ket{0}^{\rm a}\ket{\sigma_-}+t_{\rm  nc} \ket{L''}\\
\ket{1}^{\rm a}\ket{\sigma_+} &\rightarrow& r_{\rm c}\ket{1}^{\rm a}\ket{\sigma_+}+t_{\rm c} \ket{L'''} 
\end{eqnarray}

Ideally, $r_{\rm c}=1$, $r_{\rm  nc}=-1$, and $t_{\rm c}=t_{\rm  nc}=0$;  which implements an atom-photon $\sf CZ$ gate assuming the encoding $\ket{\sigma_+}=\ket{1}^{\rm p}$ and  $\ket{\sigma_-}=\ket{0}^{\rm p}$ (which is different from our scheme). As before, the initial state of the atom and the photon will have a matched and mismatched part:

\begin{align}  
 \ket{\psi}= \sqrt{1-\zeta}e^{i\theta}\ket{\psi_i,\text{mis}}+\sqrt{\zeta}\ket{\psi_i,\text{mat}}
\end{align}

where

\begin{align}  
 \ket{\psi_i}=\alpha_{\rm p}\ket{\sigma_{-}}(\alpha\ket{0}^{\rm a}+\beta\ket{1}^{\rm a})+\beta_{\rm p}\ket{\sigma_{+}}(\alpha\ket{0}^{\rm a}+\beta\ket{1}^{\rm a})
\end{align}

After the photon hits the cavity, only the matched part will experience the scattering relations above and we get:

\begin{align}  
 \ket{\psi}=\sqrt{1-\zeta}e^{i\theta}\ket{\psi_i,\text{mis}}+\sqrt{\zeta}  \bigg \{ \alpha_{\rm p}\ket{\sigma_{-},\text{mat}}(\alpha r_{\rm  nc} \ket{0}^{\rm a}+\beta r_{\rm  nc }\ket{1}^{\rm a})+\beta_{\rm p}\ket{\sigma_{+},\text{mat}}(\alpha r_{\rm  nc}\ket{0}^{\rm a}+\beta r_{\rm c}\ket{1}^{\rm a})\nonumber\\
 +\alpha_{\rm p}\alpha t_{\rm  nc}\ket{L''}+\alpha_{\rm p}\beta t_{\rm  nc}\ket{L'}+\beta_{\rm p}\alpha t_{\rm  nc}\ket{L}+\beta_{\rm p}\beta t_{\rm c}\ket{L'''} \bigg \}
\end{align}

The probability to lose the photon is

\begin{equation}  P_{\rm loss}=P(\ket{L''})+P(\ket{L'})+P(\ket{L'})+P(\ket{L'''})=\zeta \big[|t_{\rm  nc}|^2|+|\beta_{\rm p}\beta|^2(|t_{\rm c}|^2-|t_{\rm  nc}|^2)\big]
\end{equation}

To consider the case when the photon is not lost, the state gets projected into:

\begin{align}  
 \ket{\psi}=\dfrac{1}{N_l}\bigg[\sqrt{1-\zeta}e^{i\theta}\ket{\psi_i,\text{mis}}+\sqrt{\zeta}  \bigg \{ \alpha_{\rm p}\ket{\sigma_{-},\text{mat}}(\alpha r_{\rm  nc} \ket{0}^{\rm a}+\beta r_{\rm  nc }\ket{1}^{\rm a})+\beta_{\rm p}\ket{\sigma_{+},\text{mat}}(\alpha r_{\rm  nc}\ket{0}^{\rm a}+\beta r_{\rm c}\ket{1}^{\rm a}) \bigg \} \bigg]
\end{align} 

with the normalization constant $N_l=\sqrt{1-P_{\rm loss}}$. The ideal output of the $\sf CZ$ gate here is

\begin{align}  
 \ket{\psi_{\text{ideal}}}=-\big[\alpha_{\rm p}\ket{\sigma_{-}}(\alpha\ket{0}^{\rm a}+\beta\ket{1}^{\rm a})+\beta_{\rm p}\ket{\sigma_{+}}(\alpha\ket{0}^{\rm a}-\beta\ket{1}^{\rm a})\big]
\end{align}

The fidelity $|\braket{\psi_{\text{ideal}} | \psi_{\text{out}}}|^2$ is then given by

\begin{equation} 
 F= \dfrac{1}{|N_l|^2}\bigg\{ (1-\zeta ) (|\alpha_{\rm p}|^2+|\beta_{\rm p}|^2[|\alpha|^2-|\beta|^2])^2+\zeta \bigg||\alpha_{\rm p}|^2r_{\rm  nc}+|\beta_{\rm p}|^2(r_{\rm  nc}|\alpha|^2-r_{\rm c}|\beta|^2)\bigg|^2\bigg\}
\end{equation}

while the success probability is

\begin{equation} 
 P_{\rm success}=1-P_{\rm loss}=1-\zeta \big[|t_{\rm  nc}|^2|+|\beta_{\rm p}\beta|^2(|t_{\rm c}|^2-|t_{\rm  nc}|^2)\big].
\end{equation}

\rsub{
To get the average fidelity $F_{\rm avg}$, $F$ is averaged over all possible initial atom-photon product states. The average is taken over the Bloch spheres of the atom and the photon by making the substitutions
$\alpha \rightarrow \cos (\theta_1/2) e^{i \Phi_1}$, $\beta \rightarrow \sin (\theta_1/2)$, $\alpha_{\rm p} \rightarrow \cos (\theta_2/2) e^{i \Phi_2}$, and $\beta_{\rm p} \rightarrow \sin (\theta_2/2)$ in $F$:}

\begin{equation} 
\rsub{ F_{\rm avg}= \dfrac{1}{(4 \pi)^2} \int  d\theta_1 d\Phi_1  d\theta_2 d\Phi_2 \, F \sin \theta_1 \sin \theta_2  }
\end{equation}

\rsub{Where $\Phi_{1,2}$ and $\theta_{1,2}$ are the azimuthal and polar angles of the Bloch spheres, respectively.}

\subsubsection{Remote atom-atom entanglement}

In this scheme, to generate atom-atom entanglement, we start with the photon-atom-atom product state $\ket{H}\ket{0}^{\rm a}_x\ket{0}^{\rm a}_x$. The photon hits the two cavities successively, then the photon passes through a QWP. Finally, we measure its polarization, which results in an atom-atom Bell state. Again, for simplicity we assume no spatial mode mismatch. After the first $\sf CZ$ gate between $\ket{H}$ and $\ket{0}^{\rm a}_x$ we get (ignoring normalization and assuming the photon was not lost):

\begin{equation} 
\big \{ \ket{\sigma_{-}}(r_{\rm  nc}\ket{0}^{\rm a}+r_{\rm  nc}\ket{1}^{\rm a})+\ket{\sigma_{+}}(r_{\rm  nc}\ket{0}^{\rm a}+r_{\rm c}\ket{1}^{\rm a}) \big  \}\ket{0}^{\rm a}_x 
\end{equation}

where $r_{\rm c}$ and $r_{\rm  nc}$ are the reflection coefficients of the first cavity. After the $\sf CZ$ gate between the photon and the second cavity, we get

\begin{eqnarray} 
&&\big \{ \ket{\sigma_{-}}(r_{\rm  nc}\ket{0}^{\rm a}+r_{\rm  nc}\ket{1}^{\rm a})+\ket{\sigma_{+}}(r_{\rm  nc}\ket{0}^{\rm a}+r_{\rm c}\ket{1}^{\rm a}) \big  \}r'_{\rm  nc}\ket{0}^{\rm a}  \nonumber\\ 
+&&\big \{r'_{\rm  nc} \ket{\sigma_{-}}(r_{\rm  nc}\ket{0}^{\rm a}+r_{\rm  nc}\ket{1}^{\rm a})+r'_{\rm c}\ket{\sigma_{+}}(r_{\rm  nc}\ket{0}^{\rm a}+r_{\rm c}\ket{1}^{\rm a}) \big  \}\ket{1}^{\rm a}
\end{eqnarray}

where $r'_{\rm c}$ and $r'_{\rm  nc}$ are the reflection coefficients of the second cavity. After the photon passes through a QWP we get:

\begin{eqnarray} 
&&  \ket{\sigma_{-}}\big \{2 r'_{\rm  nc}r_{\rm  nc}\ket{00}^{\rm a}+( r'_{\rm  nc}r_{\rm  nc}+ r'_{\rm  nc}r_{\rm c}) \ket{10}^{\rm a}+( r'_{\rm  nc}r_{\rm  nc}+ r'_{\rm c}r_{\rm  nc})\ket{01}^{\rm a}+( r'_{\rm  nc}r_{\rm  nc}+ r'_{\rm c}r_{\rm c})\ket{11}^{\rm a} \big  \} \nonumber\\
  +&&\ket{\sigma_{+}}\big  \{( r'_{\rm  nc}r_{\rm  nc}- r'_{\rm  nc}r_{\rm c}) \ket{10}^{\rm a}+( r'_{\rm  nc}r_{\rm  nc}- r'_{\rm c}r_{\rm  nc})\ket{01}^{\rm a}+( r'_{\rm  nc}r_{\rm  nc}- r'_{\rm c}r_{\rm c})\ket{11}^{\rm a} \big  \}
\end{eqnarray}

Measuring the photon polarization gives the following atom-atom entangled states:

\begin{eqnarray}
     \ket{\sigma_{-}}&:&2 r'_{\rm  nc}r_{\rm  nc}\ket{00}^{\rm a}+( r'_{\rm  nc}r_{\rm  nc}+ r'_{\rm  nc}r_{\rm c}) \ket{10}^{\rm a}+( r'_{\rm  nc}r_{\rm  nc}+ r'_{\rm c}r_{\rm  nc})\ket{01}^{\rm a}+( r'_{\rm  nc}r_{\rm  nc}+ r'_{\rm c}r_{\rm c})\ket{11}^{\rm a} \\
     \ket{\sigma_{+}}&:&( r'_{\rm  nc}r_{\rm  nc}- r'_{\rm  nc}r_{\rm c}) \ket{10}^{\rm a}
     +( r'_{\rm  nc}r_{\rm  nc}- r'_{\rm c}r_{\rm  nc})\ket{01}^{\rm a}
     +( r'_{\rm  nc}r_{\rm  nc}- r'_{\rm c}r_{\rm c})\ket{11}^{\rm a}
\end{eqnarray}

Observe that even if the two cavities are identical we will not get the maximally entangled Bell states here. We get the maximally entangled states only if the two cavities are ideal, i.e., $r_{\rm c}=r'_{\rm c}=1$ and $r_{\rm  nc}=r'_{\rm  nc}=-1$:

\begin{eqnarray}
     \ket{\sigma_{-}}&:&\ket{00}^{\rm a}+\ket{11}^{\rm a} \\
     \ket{\sigma_{+}}&:& \ket{10}^{\rm a}+\ket{01}^{\rm a}
\end{eqnarray}

If the two cavities are not ideal, then the fidelity becomes:

\begin{eqnarray}
F(\ket{\phi_{+}})&=&\dfrac{1}{2}\dfrac{|2 r'_{\rm  nc}r_{\rm  nc}+( r'_{\rm  nc}r_{\rm  nc}+ r'_{\rm c}r_{\rm c})|^2}{|2 r'_{\rm  nc}r_{\rm  nc}|^2+|r'_{\rm  nc}r_{\rm  nc}+ r'_{\rm  nc}r_{\rm c}|^2+| r'_{\rm  nc}r_{\rm  nc}+ r'_{\rm c}r_{\rm nc}|^2+| r'_{\rm nc}r_{\rm nc}+ r'_{\rm c}r_{\rm c}|^2}\\
     F(\ket{\psi_{+}})&=&\dfrac{1}{2}\dfrac{|( r'_{\rm nc}r_{\rm nc}- r'_{\rm nc}r_{\rm c})+( r'_{\rm nc}r_{\rm nc}- r'_{\rm c}r_{\rm nc})|^2}{| r'_{\rm nc}r_{\rm nc}- r'_{\rm nc}r_{\rm c}|^2+| r'_{\rm nc}r_{\rm nc}- r'_{\rm c}r_{\rm nc}|^2+| r'_{\rm nc}r_{\rm nc}- r'_{\rm c}r_{\rm c}|^2}
\end{eqnarray}

Where $\ket{\phi_{+}}=(\ket{00}^{\rm a}+\ket{11}^{\rm a})/\sqrt{2}$ and $\ket{\psi_{+}}=(\ket{01}^{\rm a}+\ket{10}^{\rm a})/\sqrt{2}$ are the maximally entangled Bell states.

\section{Predictive power of the error model} \label{predictpower}

Here, we would like to assess how well our error model is able to estimate fidelity and success probability in actual experiments.

\subsection{Experiment 1: atom-photon entanglement}

 We compare our model (from the previous appendix) with data from experiments  by Reiserer et al. \cite{Reiserer2014}. In one of the experiments, they hit the the cavity with a single photon with the initial atom-photon state $\ket{00}^{\rm ap}_x$. Ideally, this would execute a $\sf CNOT$ gate with the atom-photon Bell state output $(\ket{00_x}^{\rm ap}+\ket{11_x}^{\rm ap})/\sqrt{2}$. The experimentally measured fidelity is 80.7\%, with the following effects leading to the infidelity. First, a non-unity mode matching efficiency of 92\% and photon loss from the cavity lead to 8(3)\% fidelity reduction. The analytic formula for fidelity from the previous Appendix is

\begin{equation} 
 F= \dfrac{1}{|N_l|^2}\bigg\{ (1-\zeta ) (|\alpha_{\rm p}|^2+|\beta_{\rm p}|^2[|\alpha|^2-|\beta|^2])^2+\zeta \bigg||\alpha_{\rm p}|^2r_{\rm nc}+|\beta_{\rm p}|^2(r_{\rm nc}|\alpha|^2-r_{\rm c}|\beta|^2)\bigg|^2\bigg\}
\end{equation}

The parameters of their experiment are \cite{Reiserer2013, Reiserer2014, Reiserer2015} $|\alpha_{\rm p}|=|\beta_{\rm p}|=|\alpha|=|\beta|=1/\sqrt{2}$, $\zeta=0.92$, $C=3$, $\Delta_{\rm c}=(\omega_{\rm p}-\omega_{\rm c})/\kappa=300 \ \rm{kHz}/( 2.5  \rm{MHz})=0.12$, $\Delta_{\rm a}=0.83\Delta_{\rm c}$ (assuming $\omega_{\rm p}-\omega_{\rm c}=\omega_{\rm p}-\omega_{\rm a}$), and $\kappa_r/\kappa\approx 2.3/2.5= 0.92$. This gives $F \approx 90 \%$. This is within two percent of the experimentally estimated fidelity reduction due to mode mismatch and photon loss. Note that our model does not take into account other effects of infidelity (atomic state preparation, readout, rotation, detector dark counts, non-ideal PBS, and multi-photon effects), which contribute to a further fidelity reduction of 10\% in the experiment. If we additionally take these effects into account, then this gives $F\approx 90-10=80\%$, in close agreement with the measured fidelity of 80.7\%. 

Next, we compare the success probability. Because of mode mismatch, 8\% of the light is totally reflected. Of the 92\% matched light that couples to the cavity, only 69\% is measured to reflected. Thus, the experimentally measured total probability that the photon is reflected back is $0.08+(0.92)(0.69)=71.5\%$. This also measures the success probability of the scheme (loss due to the photon not reflecting from the cavity leads to failure). From the previous section our formula gives:

\begin{equation} 
 P_{\rm success}=1-P_{\rm loss}=1-\zeta \big[|t_{\rm nc}|^2|+|\beta_{\rm p}\beta|^2(|t_{\rm c}|^2-|t_{\rm nc}|^2)\big]
\end{equation}

Plugging in the same parameters above, we get  $P_{\rm success}=69\%$, which is within two and a half percent of the measured value. 

\subsection{Experiment 2: atom-atom entanglement}

The second experiment we analyze is by Welte et al. \cite{Welte2018}. Here they also use the same energy scheme from the previous section. They have two atoms in a cavity. By reflecting $\ket{\sigma_+}$ off the cavity, the atoms will experience no phase shift when either one or both are coupled to the photon (i.e., when the atoms are in $\ket{11}^{\rm a},\ket{10}^{\rm a}$, or $\ket{01}^{\rm a}$), and they will experience a sign flip only when both atoms are uncoupled (i.e., $\ket{00}^{\rm a}$). The photon is always disentangled from the two atoms before and after reflection. If we have the initial atom-atom superposition state $\ket{11}_x^{\rm a}=1/2(\ket{11}^{\rm a}-\ket{10}^{\rm a}-\ket{01}^{\rm a}+\ket{00}^{\rm a})$, then scattering a $\ket{\sigma_+}$ photon creates the entangled atom-atom state $1/2(\ket{11}^{\rm a}-\ket{10}^{\rm a}-\ket{01}^{\rm a}-\ket{00}^{\rm a})$ (equivalent to a Bell state up to a global rotation). Using an error model and experimental data, the authors found that in the absence of all errors except cavity losses and finite cooperativity, the estimated fidelity (assuming the photon is detected after reflection) is $99.7\%$. They also found that spatial mode mismatch alone contributes $6\%$ reduction in fidelity (assuming everything else is ideal). The measured probability to lose the photon during reflection is  $33\%$. If we carry out the same analysis as before for this particular initial atom-atom state, we will arrive at the following equation for the fidelity of the final atom-atom Bell state:

\begin{equation} 
 F= \dfrac{1}{1-P_{\text{loss}}}\bigg\{ (1-\zeta ) \dfrac{1}{4}+\zeta \bigg|\dfrac{3 r_{\rm c}-r_{\rm nc}}{4}\bigg|^2\bigg\}
\end{equation}

with

\begin{equation} 
P_{\text{loss}}= \dfrac{\zeta}{4}(3 |t_{\rm c}|^2+|t_{\rm nc}|^2)
\end{equation}

The parameters of this experiment are   $\zeta=0.92$, $C=4$, $\Delta_{\rm c}=\Delta_{\rm a}=0$ (the authors set it to zero during their error analysis when computing the reflection coefficients), and $\kappa_r/\kappa\approx 2.29/2.5= 0.916$. Using our error model and the parameters of their experiment, the estimated fidelity (assuming the photon is detected after reflection and losses only due to cavity and finite cooperativity) is $99.96\%$, higher than the estimated fidelity of the authors by about $0.26\%$.

According to our formula and using the experimental parameters as input, the probability to lose the photon during reflection is $P_{\text{loss}}=32.3\%$, within $0.7\%$ of the experimentally measured value. 

To compute the contribution due to mode mismatch, we assume everything is perfect ($r_{\rm c}=1,r_{\rm nc}=-1,t_{\rm c}=0,t_{\rm nc}=0$), and set $\zeta=0.92$ in our equation for fidelity. This gives $F=(1-0.92)/4+0.92=0.94$. Therefore, the reduction in fidelity due to mode mismatch is $6\%$, in agreement with their analysis.

\section{\rsub{Photon loss and $\sf CZ$ gate fidelity}} \label{loss_fidelity}
\rsub{
It was shown in Sec. \ref{Atom-photon two-qubit gate} that the present scheme is able to execute an ideal $\sf Z$ gate by dissipating the error-inducing photons. However, it does not follow from this that this would lead to an ideal $\sf CZ$ gate (although it still leads to an improvement as shown in Fig. \ref{CZerror}). Generally, photon losses do not affect the fidelity of the $\sf CZ$ gate only if the photon loss probability is the same for all photonic and atomic qubit basis states (i.e., same for $|V\rangle|0\rangle^{\rm a}$, $|V\rangle|1\rangle^{\rm a}$,$|H\rangle|0\rangle^{\rm a}$, and $|H\rangle|1\rangle^{\rm a}$). On the other hand, if the photon loss probability is different for each basis state, then the fidelity of the $\sf CZ$ will not be 1 even if we perform a perfect $\sf Z$ gate. To illustrate with an example, consider an initial atom-photon equal superposition state $0.5|V\rangle(|0\rangle^{\rm a}+|1\rangle^{\rm a})+0.5|H\rangle(|0\rangle^{\rm a}+|1\rangle^{\rm a})$. The output of an ideal $\sf CZ$ acting on this state is }

\begin{equation} 
\rsub{
0.5|V\rangle(|0\rangle^{\rm a}+|1\rangle^{\rm a})+0.5|H\rangle(|0\rangle^{\rm a}-|1\rangle^{\rm a})
}
\end{equation}

\rsub{
Under the present scheme, numerical analysis shows that the output will be (using the parameters in Fig. \ref{CZerror} and ignoring spatial mode mismatch)
}

\begin{equation} 
\rsub{
0.548333|V\rangle(|0\rangle^{\rm a}+|1\rangle^{\rm a})+0.446465|H\rangle(|0\rangle^{\rm a}-|1\rangle^{\rm a})
}
\end{equation}

\rsub{
The present scheme performs the $\sf Z$ gate correctly (i.e., phase flip only for $|H\rangle|1\rangle^{\rm a}$). However, the relative amplitudes are not equal because the photon loss probabilities are different for $|V\rangle$ (which does not interact with the cavity so its probability of loss is zero, see Fig. \ref{CZ_gate}) and $|H\rangle$ (which experiences losses through cavity interactions). Therefore, one way to increase the $\sf CZ$ fidelity in this scheme further is to deliberately introduce losses in one of the MZI arms (labeled $p_2$ in Fig. \ref{CZ_gate}) to balance the relative populations of $|H\rangle$ and $|V\rangle$. This can be done by a tunable optical attenuator for example.}

\rsub{
It is worth pointing out that this relation between photon loss probabilities and $\sf CZ$ gate fidelity is not unique to the present scheme, and it applies to others, e.g., see Fig. 5(b) in Ref. \cite{JCho2005}, where the $\sf CZ$ fidelity is 1 only when the relative photon success probabilities are equal (see also Eq. (4) in that paper). 
}

\end{widetext}

\end{document}